\documentclass[sn-mathphys-num]{sn-jnl}% Math and Physical Sciences Numbered Reference Style 
\usepackage{graphicx}%
\usepackage{multirow}%
\usepackage{amsmath,amssymb,amsfonts}%
\usepackage{amsthm}%
\usepackage{mathrsfs}%
\usepackage[title]{appendix}%
\usepackage{xcolor}%
\usepackage{textcomp}%
\usepackage{manyfoot}%
\usepackage{booktabs}%
\usepackage{algorithm}%
\usepackage{algorithmicx}%
\usepackage{algpseudocode}%
\usepackage{listings}%
\usepackage{enumitem}
\usepackage{subcaption}
\theoremstyle{thmstyleone}%
\theoremstyle{thmstyletwo}%

\theoremstyle{thmstylethree}%

\begin{document}

\title[Article Title]{A Scalable Multi-Layered Blockchain Architecture
	for Enhanced EHR Sharing and Drug Supply Chain
	Management}

%%=============================================================%%
%% GivenName	-> \fnm{Joergen W.}
%% Particle	-> \spfx{van der} -> surname prefix
%% FamilyName	-> \sur{Ploeg}
%% Suffix	-> \sfx{IV}
%% \author*[1,2]{\fnm{Joergen W.} \spfx{van der} \sur{Ploeg} 
%%  \sfx{IV}}\email{iauthor@gmail.com}
%%=============================================================%%

\author[]{Reza Javan, Mehrzad Mohammadi, Mohammad Beheshti-Atashgah, and Mohammad Reza Aref}
\affil[] {The authors are with the Information Systems and Security Laboratory,
	Department of Electrical Engineering, Sharif University of Technology,
	Tehran 14588-89694, Iran.}
\affil[]{reza.javan@alum.sharif.edu, mohammadimehrzad11@gmail.com, Mohammad.beheshti@researcher.sharif.edu, aref@sharif.edu.}

%%==================================%%
%% Sample for unstructured abstract %%
%%==================================%%

\abstract{
		In recent years, the healthcare sector's transition to digital platforms has intensified concerns over data security, privacy, and scalability. Blockchain technology offers a decentralized, secure, and immutable solution to these challenges. This paper presents a scalable, multi-layered blockchain architecture for secure Electronic Health Record (EHR) sharing and drug supply chain management. The proposed framework introduces five distinct layers that enhance system performance, security, and patient-centric access control. By implementing parallelism, the system significantly increases transaction throughput and reduces network traffic. Our solution ensures data integrity, privacy, and interoperability, making it compatible with existing healthcare systems. Experimental results, conducted using the Caliper benchmark, demonstrate notable improvements in transaction throughput and reduced communication overhead. Additionally, the framework provides transparency and real-time drug supply chain monitoring, empowering decision-makers with critical insights.
}


%%\pacs[JEL Classification]{D8, H51}

%%\pacs[MSC Classification]{35A01, 65L10, 65L12, 65L20, 65L70}

\maketitle

\section{Introduction}\label{sec1}

In recent years, blockchain technology has garnered significant attention across various domains, including healthcare. This is primarily due to its potential to enhance data accuracy, security, and privacy while enabling trust among disparate stakeholders [1]. Blockchain is characterized by key features such as decentralization, persistency, immutability, and security. As healthcare services transition from offline to online modes, new security concerns emerge, including fragmented health data, interoperability challenges, data security, privacy issues, and scalability hurdles [2]-[8].

Traditional healthcare systems rely on centralized approaches for the storage and management of Electronic Health Records (EHRs). However, the frequent sharing and distribution of these records among stakeholders-such as hospitals, patients, and clinics, leads to time and cost intensive processes. Cloud-based health data management emerged as a solution for real-time data sharing, but it introduced memory-intensive data encryption requirements, particularly when transmitting data to the cloud [7]. To mitigate these issues, lightweight blockchain approaches have been developed, reducing computational and communication overhead by grouping participants into clusters and maintaining a single ledger per cluster [8]. These approaches specifically target security and privacy concerns when sharing sensitive healthcare data [9], [10].

In this paper, we propose a scalable, multi-layered blockchain framework designed to address critical challenges in healthcare, including security, privacy, scalability, data integrity, and traceability. The framework introduces a five-layer architecture that enables efficient EHR sharing and drug supply chain management. By integrating parallelism within Hyperledger Fabric, our approach optimizes transaction throughput and reduces network traffic while ensuring robust security measures and compliance with healthcare standards. Additionally, the framework leverages patient-centric access control to empower individuals in managing their health data while supporting interoperability with existing healthcare systems.

Previous works in this domain, such as "MedShard" [11], have explored blockchain-based approaches to scalability through sharding techniques, and others have proposed fine-grained access control methods for secure data sharing [12], [13]. However, these solutions face limitations when dealing with large-scale healthcare networks that require real-time data sharing and robust scalability. Our work advances the current state-of-the-art by focusing on both EHR sharing and drug supply chain management, with an emphasis on scalability and the use of parallelism for enhancing system performance.

The primary contributions of this paper are as follows: We propose a novel multi-layered blockchain architecture for secure and scalable healthcare data management, encompassing both EHR sharing and drug supply chain monitoring.
We implement parallelism in Hyperledger Fabric to significantly improve transaction throughput, reduce network traffic, and address scalability challenges.
We provide comprehensive performance evaluations using the Caliper benchmark to demonstrate the system's effectiveness in enhancing scalability.

The remainder of the paper is organized as follows: Section~\ref{sec2:Releated works} reviews related works on blockchain applications in healthcare. Section~\ref{sec3:Proposed system} presents the proposed system architecture and describes key design considerations. Section~\ref{sec4:Implementation And Testing} details the implementation and testing scenarios. Finally, Section~\ref{sec 5:Conclusion} concludes the paper with future research directions.

\section{Related works}\label{sec2:Releated works}

Blockchain technology has the potential to reshape the healthcare industry by addressing key challenges, including security, privacy, scalability, and interoperability. Numerous blockchain-based frameworks have emerged to tackle these issues, particularly in the domains of Electronic Health Record (EHR) sharing and drug supply chain management.

Several approaches have focused on improving EHR sharing by enhancing scalability, security, and interoperability. For instance, the "MedShard" framework employs sharding to boost scalability [11]. Sharding allows the blockchain to split into smaller, more manageable pieces, enabling better handling of large-scale healthcare data. Another notable work is the "Blockchain-Based Framework for Interoperable Electronic Health Records for an Improved Healthcare System" [12], which introduces fine-grained access control and ensures seamless interoperability with traditional EHR systems. This framework offers a method to restrict access to specific pieces of healthcare data, ensuring confidentiality and security.

In addition to EHR sharing, the pharmaceutical supply chain has garnered attention for its need for transparency and security. The "Healthcare Chain Network Framework for Monitoring and Verifying Pharmaceutical Supply Chain" [14] presents a comprehensive solution for tracking and verifying drugs throughout the supply chain, enhancing the integrity of the process. Moreover, the use of smart contracts has been explored in several studies. For example, "Automating Procurement Contracts in the Healthcare Supply Chain Using Blockchain Smart Contracts" [15] proposes a novel method of automating procurement processes within the healthcare supply chain, significantly reducing manual errors and improving efficiency.

Scalability remains a crucial factor when implementing blockchain in healthcare, as the volume of transactions and the size of data can overwhelm existing systems. Several models have been proposed to address scalability. One such model is the "Scalable Blockchain Model Using Off-Chain IPFS Storage for Healthcare Data Security and Privacy" [16], which leverages InterPlanetary File System (IPFS) to store healthcare data off-chain, thereby reducing the data load on the blockchain itself. Another relevant work is the "Blockchain Scalability Solved via Quintessential Parallel Multiprocessor" [17], which presents a parallel processing architecture to enhance the blockchain's scalability. This architecture highlights the importance of parallelism in achieving high throughput and efficient transaction processing.

Efforts to optimize the performance of healthcare blockchain systems also focus on lightweight architectures. "Lightweight Blockchain for Healthcare" [18] introduces a refined architecture that addresses performance bottlenecks by simplifying the blockchain's structure, allowing faster data processing. Additionally, the concept of mutable blocks has been explored in "A Novel Blockchain Architecture with Mutable Block and Immutable Transactions for Enhanced Scalability" [19], which offers an approach to scalability by maintaining block mutability while ensuring transaction immutability.

Our proposed framework builds upon these established approaches by offering a practical and well-structured method to enhance scalability in blockchain-based healthcare systems. Our work focuses on improving existing solutions through a thoughtfully designed multi-layered architecture. By leveraging parallelism within Hyperledger Fabric, we demonstrate significant performance improvements through detailed implementation and testing. This work highlights the effectiveness of architectural design and system optimization in addressing scalability challenges, ensuring security, privacy, and interoperability while supporting EHR sharing and drug supply chain management.
\section{Proposed System}\label{sec3:Proposed system}
\begin{figure}[btp]
	\centering
	\includegraphics[width=\textwidth]{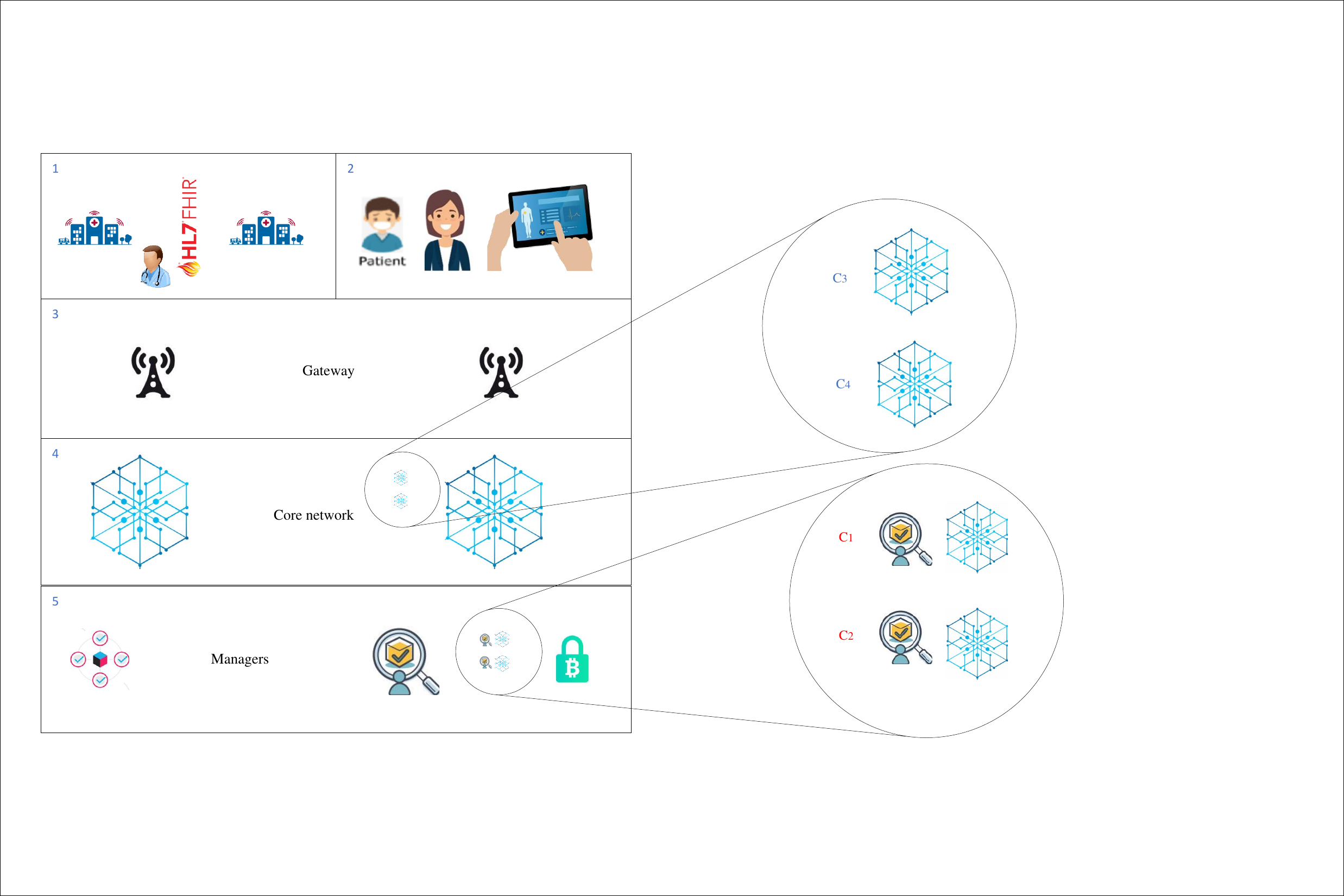}
	\caption{The proposed multi-layered system}
	\label{fig:Layers}
\end{figure}

The architecture, as shown in Figure \ref{fig:Layers}, comprises five distinct layers, each responsible for different aspects of healthcare data processing and management. These layers work cohesively to ensure smooth data flow between healthcare providers, patients, and pharmaceutical stakeholders. The layered design allows for the modular development of each component, making it adaptable to evolving healthcare needs.

Key features of the system include parallelism implemented within the Hyperledger Fabric blockchain, which enables the simultaneous processing of multiple transactions. This greatly enhances the system's scalability and throughput, ensuring it can handle a high volume of transactions typical in large-scale healthcare environments. In contrast to conventional blockchain architectures that may suffer from bottlenecks due to linear transaction processing, the parallelism in our design significantly reduces network congestion and improves efficiency.

Furthermore, the architecture adopts a patient-centric access control model. Patients are empowered to manage permissions regarding their health data, allowing them to control who can access, modify, or share their records. This approach is critical in maintaining the privacy and security of sensitive healthcare information while ensuring compliance with data protection regulations such as HIPAA and GDPR.

To ensure data integrity and interoperability, the system incorporates standardized communication protocols (such as HL7) that allow seamless integration with existing healthcare infrastructure. The use of blockchain ensures that all transactions related to EHR sharing and drug supply management are immutable and traceable, providing transparency and trust among all stakeholders involved.

\subsection{Layered Design}\label{subsec2}
The proposed multi-layered architecture is designed to distribute the responsibilities of healthcare data management across five distinct layers. Each layer is tailored to handle specific tasks such as data storage, access control, and communication between healthcare providers and patients. This modular design ensures that changes or improvements can be made to individual layers without disrupting the entire system, enhancing flexibility and scalability.

\begin{enumerate}[label=\alph*)]
\item  \textbf{Layer 1 - Medical Service Providers}

The first layer consists of medical service providers, including hospitals, doctors, clinics, and pharmacies. This layer serves as the primary interface for healthcare providers who are responsible for generating, storing, and managing patient Electronic Health Records (EHRs). Data such as patient history, diagnoses, and prescriptions are processed at this level, ensuring that only verified and authenticated users can enter sensitive data into the system.

In this layer, blockchain's decentralization feature ensures that no single provider has complete control over the data, fostering trust and transparency among the different healthcare entities involved.
\item  \textbf{Layer 2 - Medical System Users}

Layer 2 focuses on medical system users, primarily patients and their designated representatives. Patients are the core of the healthcare process, and this layer provides them with tools to control access to their health data. Using patient-centric access controls, users can specify who can read, update, or share their health records and for how long. This ensures the system aligns with regulations like HIPAA and GDPR, giving patients greater control over their personal health information.

Furthermore, this layer includes features like permission delegation, where patients can authorize healthcare professionals or family members to manage their data on their behalf. This enhances both flexibility and security in handling patient data.
\item  \textbf{Layer 3 - Gateway Layer}

The gateway layer acts as the communication bridge between medical service providers (Layer 1) and medical system users (Layer 2) to Network Core Layer (Layer 4). It processes requests from both sides, ensuring that data flows securely and efficiently between them. This layer is essential for maintaining real-time updates and ensuring that patient data is only accessed by authorized individuals. The gateway layer also ensures that all transactions are logged on the blockchain, creating a verifiable record of data exchanges.

The use of smart contracts in this layer automates the process of validating and executing transactions, such as granting data access, updating health records, or transferring permissions, thus reducing manual intervention and potential human error.
\item \textbf{Layer 4 - Network Core Layer}

The core of the blockchain network resides in Layer 4, which consists of the Hyperledger Fabric nodes. These nodes are responsible for executing and validating all the transactions that occur within the network. The parallelism implemented at this level enables the network to handle a large number of transactions simultaneously, thereby improving scalability and throughput.

Layer 4 also ensures data integrity and immutability, as each transaction is cryptographically secured and stored across multiple nodes, preventing any unauthorized alterations to the data. This layer plays a critical role in maintaining the trust and security that are fundamental to the healthcare blockchain system.
\item \textbf{Layer 5 - Controller Layer}\\
The final layer, the Controller Layer, is responsible for monitoring the execution of system policies and ensuring compliance with predefined rules. This layer consists of two major units:

The Health File Control Unit (c1), which manages patient health records, including their creation, modification, and sharing.
The Medicine Supply Control Unit (c2), which tracks the lifecycle of pharmaceuticals from production to dispensation, ensuring that each transaction involving medicine is recorded and traceable.

This layer also facilitates real-time drug supply monitoring, providing stakeholders with insights into drug availability, distribution, and potential shortages. It prevents counterfeit drugs from entering the supply chain by recording every transaction, from manufacturing to patient use, on the blockchain.
\end{enumerate}

The cryptographic security of the system ensures the confidentiality, integrity, and non-repudiation of healthcare data.
\subsubsection{Key System Transactions}\label{subsubsec2}
\begin{figure}[btp]
	\centering
	\begin{subfigure}[tb]{0.3\textwidth}
		\label{fig: Transaction 1}
		\centering
		\includegraphics[width=0.85\textwidth]{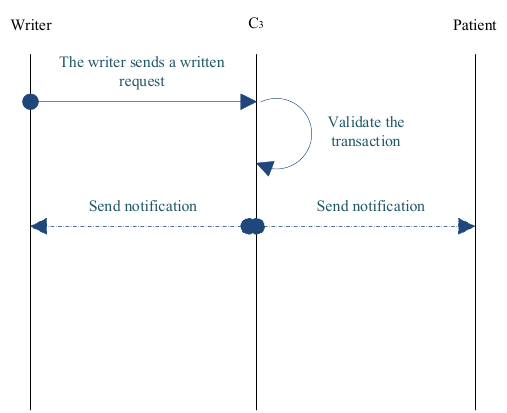}
		\caption{\footnotesize Record Health File Abstract}
	\end{subfigure}
	\hfill
	\begin{subfigure}[bt]{0.3\textwidth}
		\label{fig:Transaction 2}
		\centering
		\includegraphics[ width=\textwidth]{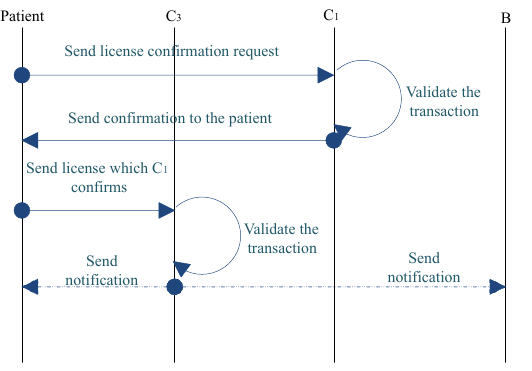}
		\caption{\footnotesize Grant Access Permissions}
	\end{subfigure}
	\hfill
	\begin{subfigure}[bt]{0.3\textwidth}
		\label{fig:Transaction 3}
		\centering
		\includegraphics[width=0.8\textwidth]{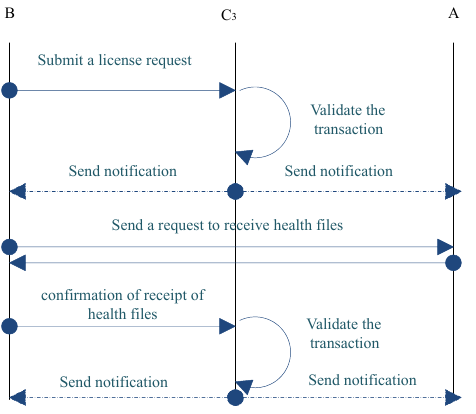} % Change the image path
		\caption{\footnotesize Health record sharing}
	\end{subfigure}
	\newline
	\newline
	\begin{subfigure}[tb]{0.3\textwidth}
		\label{fig:Transaction 4}
		\centering
		\includegraphics[width=\textwidth]{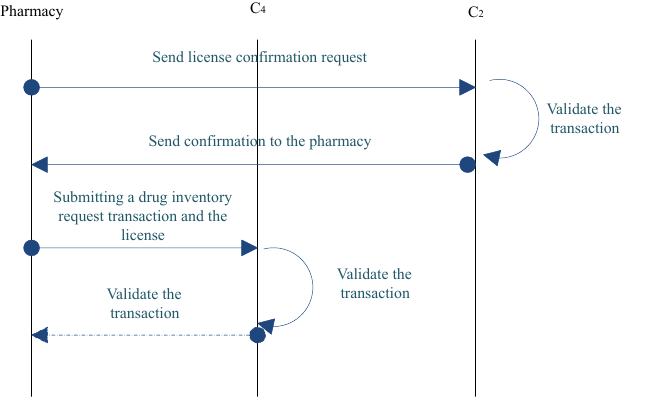} % Change the image path
		\caption{\footnotesize Register Medicine Receipt}
	\end{subfigure}
	\qquad
	\begin{subfigure}[tb]{0.3\textwidth}
		\label{fig:Transaction 5}
		\centering
		\includegraphics[width=\textwidth]{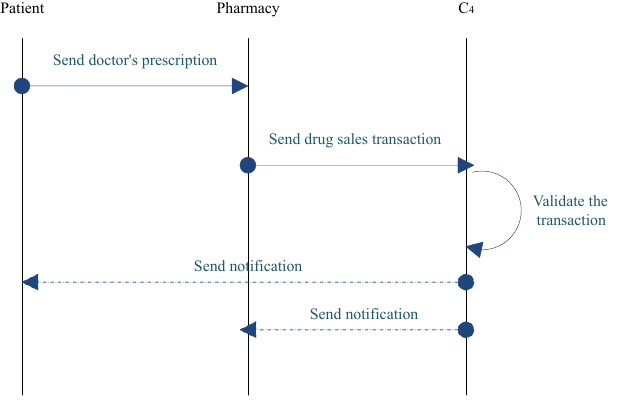}
		\caption{\footnotesize Sell Medicine to Patient}
	\end{subfigure}
	\caption{Key System Transactions}
\end{figure}
The proposed architecture enables five essential transactions designed to manage both healthcare records and the pharmaceutical supply chain. Each transaction is carefully structured to ensure data privacy, security, and scalability, all while maintaining transparency and traceability across the healthcare ecosystem.

\begin{enumerate}
\item \textbf{Record Health File Abstract}\\
The first transaction involves the creation of a health file abstract within the blockchain. Due to the vast size of Electronic Health Records (EHRs), storing raw data directly on the blockchain is inefficient and impractical. Instead, our system stores a hash of the health record, along with critical metadata such as the record's creation timestamp, patient ID, and author ID (doctor, hospital, or clinic). This ensures that the record is immutable and traceable.

Once the health record abstract is created, it can be shared securely among the different parties involved in the patient's care, without exposing the entire medical history to unauthorized individuals. This transaction provides both data integrity and privacy, as only authorized individuals can access or modify the actual health records.
\item \textbf{Grant Access Permissions}

In this transaction, patients can grant specific access permissions to healthcare providers, allowing them to read, write, update, or delete particular health records. The patient can also delegate power of attorney to trusted representatives, enabling them to manage permissions on their behalf.

The system enforces access control via smart contracts, ensuring that only those with the appropriate permissions can access sensitive data. Permissions are logged as transactions within the blockchain, providing a transparent, immutable record of who has access to what data and for how long.

Additionally, patients can revoke access at any time, ensuring full control over their health information. This transaction empowers patients while ensuring compliance with privacy regulations like HIPAA and GDPR.
\item  \textbf{Health Record Sharing}

Health record sharing occurs when a patient or healthcare provider needs to transfer specific health information to another authorized entity. The system follows the Health Level Seven (HL7) standard to ensure secure and standardized communication between different healthcare providers.

In this transaction, the requesting party submits a formal request for health information, which is then validated by the blockchain nodes. Once the request is approved, the health record abstract (along with the actual data, if necessary) is shared through a secure channel. Both the sender and the receiver must sign the transaction to confirm the transfer of information, ensuring non-repudiation and authenticity.

This transaction ensures that health data is shared securely and efficiently, preventing unauthorized access while maintaining data integrity.

\item \textbf{Register Medicine Receipt}

The fourth key transaction focuses on the pharmaceutical supply chain, specifically when a pharmacy receives a shipment of drugs. The pharmacy initiates a drug registration transaction, which is signed and validated by the system's nodes. The transaction contains essential information such as the drug's quick response (QR) code, name, expiration date, and the drug's tracking number.

This transaction is crucial for ensuring the traceability and authenticity of medications, preventing counterfeit drugs from entering the supply chain. Once the transaction is validated, it is logged on the blockchain, and the pharmacy gains access to sell the drug to patients.
\item \textbf{Sell Medicine to Patient}

In this final transaction, the system records the sale of medication to a patient. When a patient presents a prescription, the pharmacy verifies it through the blockchain. The prescription contains a signed hash that validates the patient's identity and the authenticity of the prescribed medication.

The pharmacy logs the sale as a blockchain transaction, including key details such as the patient ID, prescription ID, drug information, and transaction timestamp. This ensures transparency and non-repudiation of the sale, as both the pharmacy and patient can verify the transaction at any time.
\end{enumerate}

\subsection{Scalability via Parallelism}
	Traditional blockchain architectures typically process transactions in a sequential manner, which can lead to bottlenecks as transaction volume increases. This is particularly problematic in healthcare, where data flows from various sources (hospitals, pharmacies, patients, and regulatory bodies) at a high rate. To mitigate these bottlenecks, our system implements a parallel processing architecture where different types of transactions (e.g., EHR sharing, medication dispensing, access control) are processed concurrently across multiple nodes in the network. Each node in the Hyperledger Fabric network is designed to handle a specific subset of transactions. By distributing the workload across multiple nodes and executing transactions in parallel, the system avoids congestion and achieves higher throughput. This design is particularly beneficial for large-scale healthcare networks, where thousands of transactions can occur simultaneously, such as during peak hospital hours or nationwide medical emergencies.
\section{Implementation And Testing}\label{sec4:Implementation And Testing}
	The proposed multi-layered blockchain architecture was implemented using Hyperledger Fabric, a permissioned blockchain platform known for its modularity, privacy, and scalability features, making it an ideal choice for healthcare applications. Hyperledger Fabric is flexible architecture allowed us to tailor the system to meet the specific needs of Electronic Health Record (EHR) sharing and drug supply chain management while enabling parallel transaction processing.
	
\begin{enumerate}
	\item \textbf{Network Setup}:
	The network was designed to simulate a real-world healthcare environment, consisting of multiple nodes that represent different stakeholders, such as hospitals, clinics, pharmacies, and drug manufacturers. The implementation was deployed using Docker containers, which enabled the creation of isolated, lightweight environments for each node. The containers were orchestrated to form a cohesive blockchain network, ensuring that each participant could communicate securely and efficiently.
	
	The network consisted of:
	\begin{itemize}

		\item Two organizations representing different healthcare entities.
		\item Multiple peer nodes within each organization, responsible for endorsing and validating transactions.
		\item Ordering nodes to ensure the correct sequencing and confirmation of transactions across the network.
		\item Licensing nodes responsible for managing permissions and access control within the system.
	\end{itemize}
	All participants in the network interact through predefined smart contracts, also known as chaincode in Hyperledger Fabric, which manage the logic for transactions such as health record access and drug registration.
	\item \textbf{Parallelism Implementation}:
		To address the challenge of scalability, parallelism was introduced into the system's transaction processing. Each organization within the network operated on a separate blockchain, handling its own subset of transactions independently. This parallel transaction processing significantly improved the system's overall throughput by distributing the workload across multiple nodes.
		
		To facilitate communication between organizations or across different healthcare institutions, secure inter-organizational channels were established. These channels allowed the system to manage data sharing and drug supply chain information without overloading the main blockchain, ensuring efficient cross-chain communication when necessary.
	\item \textbf{Testing Tools and Frameworks}:
		We utilized Hyperledger Caliper, an open-source blockchain benchmarking tool, to assess the performance of the implemented system. Caliper allowed us to conduct various tests to measure key performance indicators, such as:
		\begin{itemize}
			\item Transaction throughput: The number of transactions the system can process per second (TPS).
			\item Network traffic: The amount of data transferred between nodes during transaction processing.
			\item Resource utilization: Metrics such as CPU and RAM usage during peak load and normal operation.
		\end{itemize}
		
		All tests were performed on a local setup using an Ubuntu-based environment. The network and peer nodes were deployed within Docker containers, and each container was configured to simulate real-world healthcare network conditions. The hardware used for the implementation included a Core i7 CPU with 12 GB of RAM, providing sufficient computational power to test the scalability and parallelism features of the system.
	
\end{enumerate}
\subsection{Detailed Testing Scenarios}
To evaluate the performance and scalability of our proposed system, we implemented five distinct scenarios. Each scenario was tested using two different network architectures, as depicted in Figure 3. The first two scenarios focused on drug sales registration, while the latter three were designed to test the system's ability to retrieve information from the blockchain ledgers. Across all scenarios, we utilized 20 clients to simulate real-world conditions and tested the system under maximum load to stress-test its capabilities.
\begin{figure}[bt]
	\centering
	\begin{subfigure}{0.4\textwidth}
		\centering
		\includegraphics[width=\textwidth]{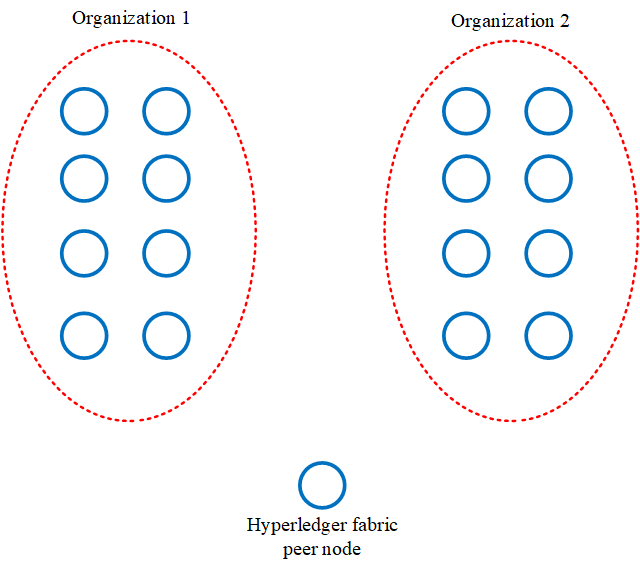}
		\caption{First structure}
		\label{fig:The fisrt structure}
	\end{subfigure} 
	\hfill
	\begin{subfigure}{0.4\textwidth}
		\centering
		\includegraphics[width=\textwidth]{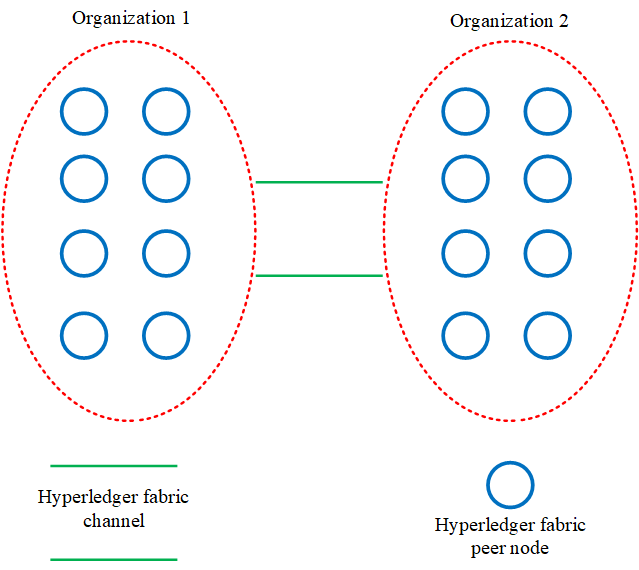}
		\caption{Second structure}
		\label{fig:The second structure}
	\end{subfigure}
	\caption{Test structures}
\end{figure}
\begin{enumerate}
\item \textbf{Scenario 1: Parallel Drug Sales Registration}

In the first scenario, we tested the system's ability to process drug sales transactions using two independent parallel blockchains (as illustrated in Figure \ref{fig:The fisrt structure}).  Each blockchain was composed of eight peer nodes, totaling 16 nodes in the network. These two parallel blockchains were designed to operate independently, with each blockchain processing 5,000 drug sales transactions, resulting in a total of 10,000 transactions being processed across the network. Each transaction involved the sale of a drug and its corresponding ledger entry, ensuring that the blockchain accurately reflected the transaction history for each drug.
\newline
\textbullet\ \textbf{Why Simulated}: This scenario tests the system's ability to process drug sales transactions in a fully parallel setup, with two independent blockchains.
\newline
\textbullet\ \textbf{Key Difference}: Two independent blockchains each process 5,000 transactions, totaling 10,000 drug sales.

\item \textbf{Scenario 2: Integrated Drug Sales Registration}

The second scenario was similar to the first but utilized the network architecture depicted in Figure \ref{fig:The second structure}. In this setup, the two blockchains were not independent but instead collaborated through a shared communication channel within the Hyperledger Fabric network. This channel allowed the two blockchains to interact and jointly process the 10,000 drug sales transactions, as opposed to working in isolation. This scenario tested the system's ability to handle collaborative processing and the effect of shared resources on transaction throughput and network traffic.
\newline
\textbullet\ \textbf{Why Simulated}: To evaluate how a collaborative blockchain setup, using a shared communication channel, affects performance in comparison to independent blockchains.
\newline
\textbullet\ \textbf{Key Difference}: Two blockchains work together through a shared channel to process 10,000 transactions, focusing on how interdependence impacts performance.

\item \textbf{Scenario 3: Parallel Information Retrieval (500 Drugs)}

The third scenario focused on retrieving drug information from the blockchain ledgers. In this case, we employed the same parallel blockchain setup from Scenario 1 (Figure \ref{fig:The fisrt structure}), with two independent blockchains and eight peer nodes in each. Each blockchain contained information for 500 drugs. The goal of this scenario was to evaluate how efficiently the system could retrieve data from two parallel blockchains and measure the performance impact of parallel data handling. By analyzing the results, we could determine whether parallelism provided a tangible benefit in information retrieval tasks.
\\
\textbullet\ \textbf{Why Simulated}: This scenario assesses how efficiently the system retrieves information from two parallel blockchains, each containing 500 drug records.
\\
\textbullet\ \textbf{Key Difference}: Parallel blockchains each hold data for 500 drugs, focusing on the benefits of parallelism in data retrieval tasks.
\item \textbf{Scenario 4: Non-Parallel Information Retrieval (1,000 Drugs)}

The fourth scenario was designed to explore the impact of non-parallel information retrieval. The setup was similar to Scenario 3 but increased the number of drugs stored in each ledger to 1,000. This allowed us to test the system's performance when handling larger datasets without utilizing parallelism. By comparing the results from Scenario 3 (parallel) and Scenario 4 (non-parallel), we aimed to identify the potential performance bottlenecks that occur when parallelization is not employed in data retrieval operations.
\\
\textbullet\ \textbf{Why Simulated}: This scenario tests the impact of non-parallel information retrieval with larger datasets, revealing performance bottlenecks when parallelism isn't employed.
\\
\textbullet\ \textbf{Key Difference}: A non-parallel structure retrieves data from 1,000 drugs in each ledger, highlighting the effects of non-parallel data handling.
\item \textbf{Scenario 5: Integrated Information Retrieval (1,000 Drugs)}

In the fifth and final scenario, the two blockchains were merged into a single channel, as shown in Figure \ref{fig:The second structure}, and each node stored information for 1,000 drugs. This scenario tested the system's ability to retrieve data from a single integrated blockchain channel, where all nodes share the same communication resources. The purpose was to assess how the system performed under a fully integrated structure compared to the parallel structures used in Scenarios 3 and 4.
\\
\textbullet\ \textbf{Why Simulated}: To compare the performance of a fully integrated blockchain channel, where all nodes share communication resources, with parallel and non-parallel structures.
\\
\textbullet\ \textbf{Key Difference}: A single channel is used to retrieve data for 1,000 drugs across all nodes, focusing on the performance in an integrated network structure. \vspace{0.25cm}

Across all scenarios, the network was subjected to the maximum load by ensuring that each client was continuously performing operations, the Caliper benchmark tool was used to evaluate key performance metrics.
\end{enumerate}
\subsection{Performance Metrics}
To assess the effectiveness and scalability of the proposed blockchain architecture, several key performance metrics were utilized. These metrics provide a comprehensive understanding of the system's operational efficiency, resource utilization, and scalability in handling healthcare data transactions. Below is a brief explanation of each metric:
	\begin{table}[htbp]
			\centering
			\caption{Scenario 1: Parallel Drug Sales Registration}
			\begin{tabular}{|c|c|c|c|c|}
				\toprule
				\textbf{Nodes} & \textbf{CPU \%} & \textbf{Memory} & \textbf{Traffic Input} & \textbf{Traffic Output} \\ \midrule
				p0o1           & 1.23\%          & 128.8 Mib       & 112 MB                 & 74 MB                   \\ \midrule
				p1o1           & 2.65\%          & 133.5 Mib       & 116 MB                 & 86 MB                   \\ \midrule
				p2o1           & 2.54\%          & 145.6 Mib       & 123 MB                 & 119 MB                  \\ \midrule
				p3o1           & 2.36\%          & 143.5 Mib       & 121 MB                 & 117 MB                 \\  \midrule
				p4o1           & 2.35\%          & 145.7 Mib       & 119 MB                 & 113 MB                  \\ \midrule
				p5o1           & 2.41\%          & 145.6 Mib       & 119 MB                 & 119 MB                 \\ \midrule
				p6o1           & 2.35\%          & 134.2 Mib       & 121 MB                 & 116 MB                 \\ \midrule
				p7o1           & 2.44\%          & 137 Mib         & 123 MB                 & 123 MB                 \\ \midrule
				couchdb0       & 1.73\%          & 61.7 Mib        & 1 MB                   & 1 MB                   \\ \midrule
				couchdb1       & 1.73\%          & 61.7 Mib        & 1 MB                   & 1 MB                   \\ \midrule
				ordering node  & 1.78\%          & 78 Mib          & 25 MB                  & 149 MB				 \\	
				\botrule
			\end{tabular}
			\label{table:Scenario 1}
	\end{table}
	\begin{table}[htbp]
			\centering
			\caption{Scenario 2: Integrated Drug Sales Registration}
			\begin{tabular}{|c|c|c|c|c|}
				\toprule
				\textbf{Nodes} & \textbf{CPU \%} & \textbf{Memory} & \textbf{Traffic Input} & \textbf{Traffic Output} \\ \midrule
				p0o1           & 1.23\%          & 128.8 Mib       & 112 MB                 & 74 MB                   \\ \midrule
				p1o1           & 2.65\%          & 133.5 Mib       & 116 MB                 & 86 MB                   \\ \midrule
				p2o1           & 2.54\%          & 145.6 Mib       & 123 MB                 & 119 MB                  \\ \midrule
				p3o1           & 2.36\%          & 143.5 Mib       & 121 MB                 & 117 MB                  \\ \midrule
				p4o1           & 2.35\%          & 145.7 Mib       & 119 MB                 & 113 MB                  \\ \midrule
				p5o1           & 2.41\%          & 145.6 Mib       & 119 MB                 & 119 MB                  \\ \midrule
				p6o1           & 2.35\%          & 134.2 Mib       & 121 MB                 & 116 MB                  \\ \midrule
				p7o1           & 2.44\%          & 137 Mib         & 123 MB                 & 123 MB                  \\ \midrule
				p0o2           & 1.23\%          & 128.8 Mib       & 112 MB                 & 74 MB                   \\ \midrule
				p1o2           & 2.65\%          & 133.5 Mib       & 116 MB                 & 86 MB                   \\ \midrule
				p2o2           & 2.54\%          & 145.6 Mib       & 123 MB                 & 119 MB                  \\ \midrule
				p3o2           & 2.36\%          & 143.5 Mib       & 121 MB                 & 117 MB                  \\ \midrule
				p4o2           & 2.35\%          & 145.7 Mib       & 119 MB                 & 113 MB                  \\ \midrule
				p5o2           & 2.41\%          & 145.6 Mib       & 119 MB                 & 119 MB                  \\ \midrule
				p6o2           & 2.35\%          & 134.2 Mib       & 121 MB                 & 116 MB                  \\ \midrule
				p7o2           & 2.44\%          & 137 Mib         & 123 MB                 & 123 MB                  \\ \midrule
				couchdb0       & 1.73\%          & 61.7 Mib        & 1 MB                   & 1 MB                    \\ \midrule
				couchdb1       & 1.73\%          & 61.7 Mib        & 1 MB                   & 1 MB                    \\ \midrule
				ordering node  & 1.78\%          & 78 Mib          & 25 MB                  & 149 MB   					\\ 
				\botrule
			\end{tabular}
			\label{table:Scenario 2}
	\end{table}
		
	\begin{table}[htbp]
			\centering
			\caption{Scenario 3: Parallel Information Retrieval (500 Drugs)}
		    \begin{tabular}{|c|c|c|c|c|}
			\hline
			
			\toprule
			\textbf{Nodes} & \textbf{CPU \%} & \textbf{Memory} & \textbf{Traffic Input} & \textbf{Traffic Output} \\ \midrule
			p0o1           & 2.21\%          & 112.8 Mib       & 93 MB                  & 106 MB                  \\ \midrule
			p1o1           & 0.82\%          & 101.4 Mib       & 35 MB                  & 104 MB                  \\ \midrule
			p2o1           & 0.85\%          & 101.5 Mib       & 35 MB                  & 104 MB                  \\ \midrule
			p3o1           & 0.83\%          & 96.7 Mib        & 35 MB                  & 104 MB                  \\ \midrule
			p4o1           & 0.85\%          & 95.1 Mib        & 35 MB                  & 104 MB                  \\ \midrule
			p5o1           & 0.85\%          & 101.2 Mib       & 35 MB                  & 104 MB                  \\ \midrule
			p6o1           & 0.83\%          & 98.9 Mib        & 35 MB                  & 104 MB                  \\ \midrule
			p7o1           & 0.81\%          & 96.1 Mib        & 35 MB                  & 104 MB                  \\ \midrule
			p0o2           & 2.21\%          & 112.8 Mib       & 93 MB                  & 106 MB                  \\ \midrule
			p1o2           & 0.82\%          & 101.4 Mib       & 35 MB                  & 104 MB                  \\ \midrule
			p2o2           & 0.85\%          & 101.5 Mib       & 35 MB                  & 104 MB                  \\ \midrule
			p3o2           & 0.83\%          & 96.7 Mib        & 35 MB                  & 104 MB                  \\ \midrule
			p4o2           & 0.85\%          & 95.1 Mib        & 35 MB                  & 104 MB                  \\ \midrule
			p5o2           & 0.85\%          & 101.2 Mib       & 35 MB                  & 104 MB                  \\ \midrule
			p6o2           & 0.83\%          & 98.9 Mib        & 35 MB                  & 104 MB                  \\ \midrule
			p7o2           & 0.81\%          & 96.1 Mib        & 35 MB                  & 104 MB                  \\ \midrule
			couchdb0       & 8.80\%          & 63.2 Mib        & 2 MB                   & 100 MB                  \\ \midrule
			couchdb1       & 8.80\%          & 63.2 Mib        & 2 MB                   & 100 MB                  \\ \midrule
			ordering node  & 0.01\%          & 22.3 Mib        & 0 MB                     & 0 MB                      \\ 
			\botrule
		\end{tabular}
		\label{table:Scenario 3}
	\end{table}
	\begin{table}[htbp]
		\centering
		\caption{Scenario 4: Non-Parallel Information Retrieval (1,000 Drugs)}
		\begin{tabular}{|c|c|c|c|c|}
			\toprule
			\textbf{Nodes}       & \textbf{CPU \%} & \textbf{Memory} & \textbf{Traffic Input} & \textbf{Traffic Output} \\ \midrule
			p0o1          & 3.31\%          & 105 Mib        & 219 MB                 & 187 MB                  \\ \midrule
			p1o1          & 0.87\%          & 100 Mib        & 60 MB                  & 183 MB                  \\ \midrule
			p2o1          & 0.89\%          & 98.46 Mib      & 61 MB                  & 183 MB                  \\ \midrule
			p3o1          & 0.91\%          & 110.57 Mib     & 61 MB                  & 183 MB                  \\ \midrule
			p4o1          & 0.89\%          & 103.89 Mib     & 61 MB                  & 183 MB                  \\ \midrule
			p5o1          & 0.94\%          & 107 Mib        & 61 MB                  & 183 MB                  \\ \midrule
			p6o1          & 0.89\%          & 98.11 Mib      & 61 MB                  & 183 MB                  \\ \midrule
			p7o1          & 0.91\%          & 106.25 Mib     & 61 MB                  & 183 MB                  \\ \midrule
			p0o2          & 3.31\%          & 105 Mib        & 219 MB                 & 187 MB                  \\ \midrule
			p1o2          & 0.87\%          & 100 Mib        & 60 MB                  & 183 MB                  \\ \midrule
			p2o2          & 0.89\%          & 98.46 Mib      & 61 MB                  & 183 MB                  \\ \midrule
			p3o2          & 0.91\%          & 110.57 Mib     & 61 MB                  & 183 MB                  \\ \midrule
			p4o2          & 0.89\%          & 103.89 Mib     & 61 MB                  & 183 MB                  \\ \midrule
			p5o2          & 0.94\%          & 107 Mib        & 61 MB                  & 183 MB                  \\ \midrule
			p6o2          & 0.89\%          & 98.11 Mib      & 61 MB                  & 183 MB                  \\ \midrule
			p7o2          & 0.91\%          & 106.25 Mib     & 61 MB                  & 183 MB                  \\ \midrule
			couchdb0      & 12.58\%         & 66 Mib         & 4 MB                   & 4 MB                    \\ \midrule
			couchdb1      & 12.58\%         & 66 Mib         & 4 MB                   & 4 MB                    \\ \midrule
			ordering node & 0.00\%          & 21 Mib         & 0 MB                     & 165 MB                  \\ 
			\botrule
		\end{tabular}
		\label{table:Scenario 4}
	\end{table}
	\begin{table}[htbp]
		\centering
		\caption{Scenario 5: Integrated Information Retrieval (1,000 Drugs)}
		\begin{tabular}{|c|c|c|c|c|}
			\toprule
			\textbf{Nodes}       & \textbf{CPU \%} & \textbf{Memory}    & \textbf{Traffic Input} & \textbf{Traffic Output} \\ \midrule
			p0o1          & 2.76\%          & 114.99 Mib    & 489 MB                 & 420 MB                  \\ \midrule
			p1o1          & 1.00\%          & 119 Mib       & 136 MB                 & 407 MB                  \\ \midrule
			p2o1          & 1.00\%          & 126 Mib       & 136 MB                 & 406 MB                  \\ \midrule
			p3o1          & 1.00\%          & 110 Mib       & 136 MB                 & 407 MB                  \\ \midrule
			p4o1          & 1.00\%          & 110 Mib       & 136 MB                 & 407 MB                  \\ \midrule
			p5o1          & 1.00\%          & 110 Mib       & 136 MB                 & 406 MB                  \\ \midrule
			p6o1          & 1.00\%          & 110 Mib       & 136 MB                 & 408 MB                  \\ \midrule
			p7o1          & 1.00\%          & 110 Mib       & 136 MB                 & 407 MB                  \\ \midrule
			p0o2          & 0.31\%          & 82.35 Mib     & 4 MB                   & 2 MB                    \\ \midrule
			p1o2          & 0.35\%          & 119 Mib       & 4 MB                   & 4 MB                    \\ \midrule
			p2o2          & 0.32\%          & 85 Mib        & 4 MB                   & 4 MB                    \\ \midrule
			p3o2          & 0.34\%          & 103.6 Mib     & 4 MB                   & 4 MB                    \\ \midrule
			p4o2          & 0.28\%          & 105.7 Mib     & 4 MB                   & 4 MB                    \\ \midrule
			p5o2          & 0.36\%          & 83.4 Mib      & 4 MB                   & 4 MB                    \\ \midrule
			p6o2          & 0.34\%          & 98.1 Mib      & 4 MB                   & 4 MB                    \\ \midrule
			p7o2          & 0.35\%          & 81 Mib        & 4 MB                   & 4 MB                    \\ \midrule
			couchdb0      & 11.60\%         & 57 Mib        & 9 MB                   & 352 MB                  \\ \midrule
			couchdb1      & 0.25\%          & 66 Mib        & 1 MB                   & 1 MB                    \\ \midrule
			orderer       & 0.00\%          & 22.47 Mib     & 0 MB                   & 0 MB                    \\ 
			\botrule
		\end{tabular}
		\label{table:Scenario 5}
	\end{table}
	
\begin{table}[htbp]
	\centering
	\caption{Performance Metrics Evaluation of Five Scenarios}
	\begin{tabular}{|c|c|c|c|c|c|}
		\toprule
		\textbf{Metric} & \textbf{Scenario 1} & \textbf{Scenario 2} & \textbf{Scenario 3} & \textbf{Scenario 4} & \textbf{Scenario 5} \\ \midrule
		
		\textbf{CPU \%} & 42\% & 32\% & 34\% & 44\% & 24\% \\ \midrule
		
		\textbf{Memory (MiB)} & 2429 & 2558 & 1756 & 1812 & 1482 \\ \midrule
		
		\textbf{Input Traffic (MB)} & 1935 & 3013 & 681 & 1298 & 1428 \\ \midrule
		
		\textbf{Output Traffic (MB)} & 1859 & 3070 & 1868 & 3267 & 3651 \\ \midrule
		
		\textbf{Total Traffic (MB)} & 3794 & 6083 & 2549 & 4565 & 5133 \\ \midrule
		
		\textbf{Throughput (Tps)} & 129 & 37 & 118 & 78 & 36 \\ 
		\botrule
	\end{tabular}
	\label{table:performance_metrics}
\end{table}

\begin{enumerate}
\item \textbf{Transaction Throughput (TPS)}:
This metric measures the number of transactions the system can process per second (TPS). In healthcare applications, high throughput is essential for ensuring that large volumes of data, such as Electronic Health Records (EHR) updates and pharmaceutical transactions, are processed in a timely manner.

\item \textbf{Resource Utilization (CPU and Memory)}:
Efficient use of computational resources is critical for ensuring the system can scale without overwhelming the hardware. This metric tracks the percentage of CPU usage and memory consumption during transaction processing, particularly under heavy load. Low resource utilization indicates that the system is capable of handling more transactions without additional hardware requirements.

\item \textbf{Network Traffic (Input/Output Traffic)}:
In a decentralized system like blockchain, nodes communicate frequently to validate and propagate transactions. This metric monitors the amount of input and output data traffic between nodes, providing insights into the system's efficiency in managing network bandwidth.\\
\end{enumerate}
\subsection{Comparison of Parallel and Non-Parallel Architectures}
\begin{figure}[htbp]
	\centering
	
	\begin{subfigure}{0.3\textwidth}
		\centering
		\includegraphics[width=\textwidth]{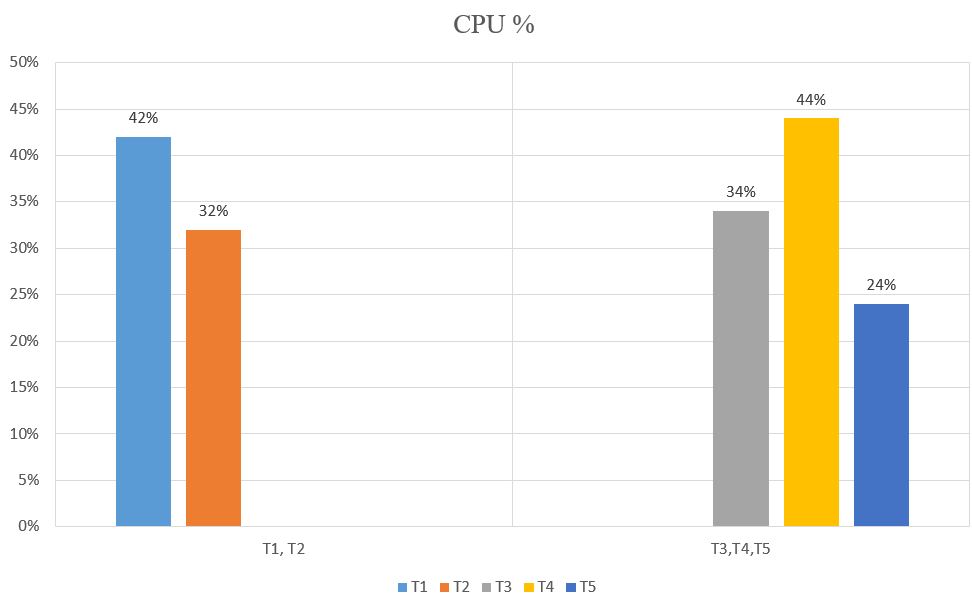}
		\caption{CPU}
	\end{subfigure}
	\hfill
	\begin{subfigure}{0.3\textwidth}
		\centering
		\includegraphics[width=\textwidth]{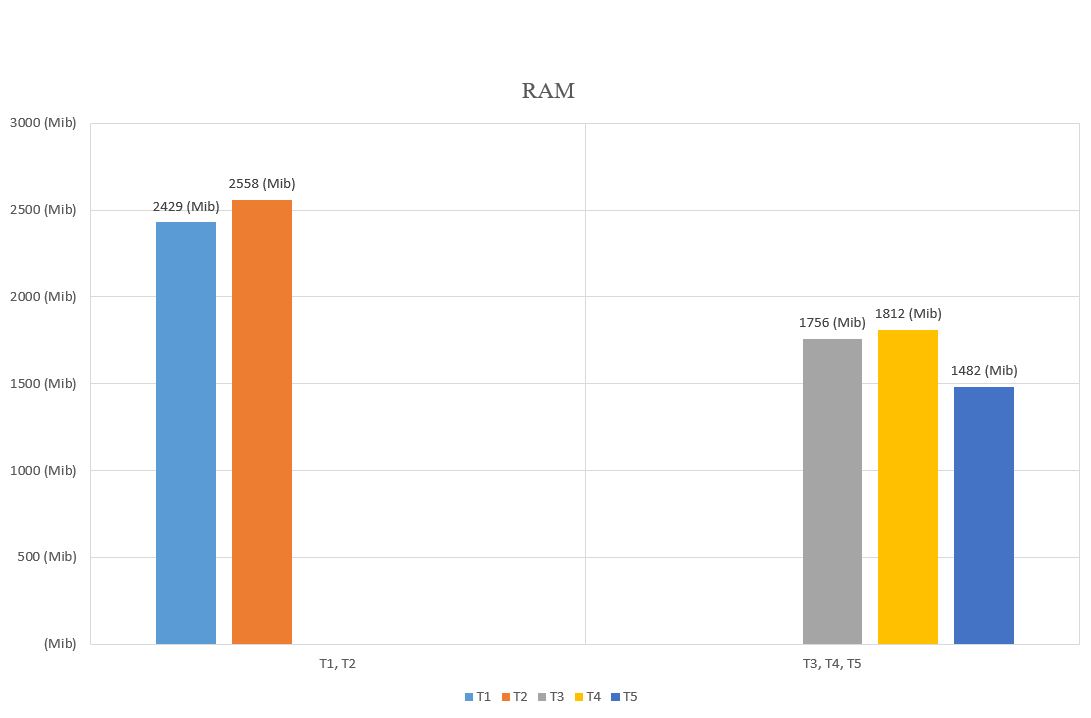}
		\caption{RAM}
	\end{subfigure}
	\hfill
	\begin{subfigure}{0.3\textwidth}
		\centering
		\includegraphics[width=\textwidth]{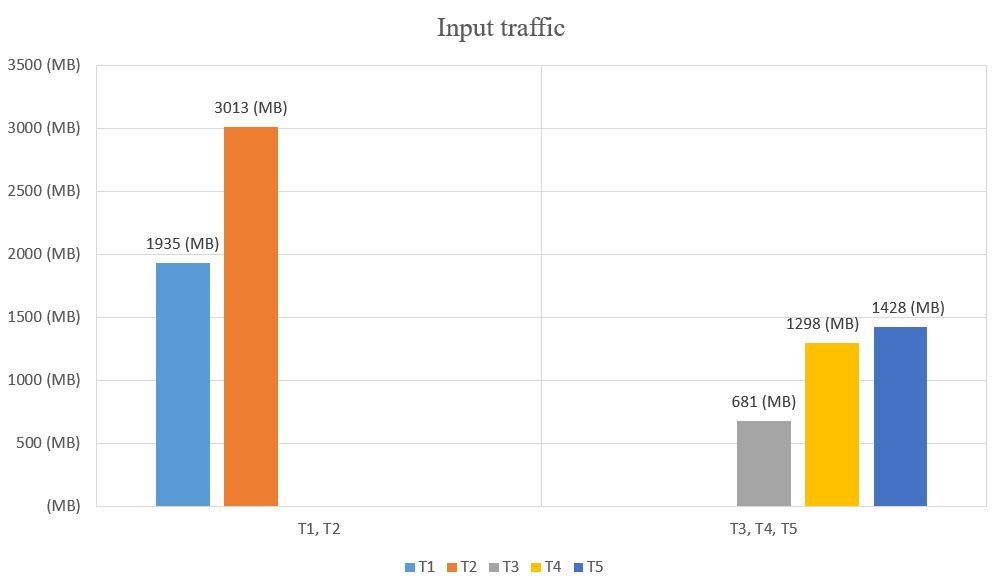}
		\caption{Input traffic}
	\end{subfigure}
	\\
	\begin{subfigure}{0.3\textwidth}
		\centering
		\includegraphics[width=\textwidth]{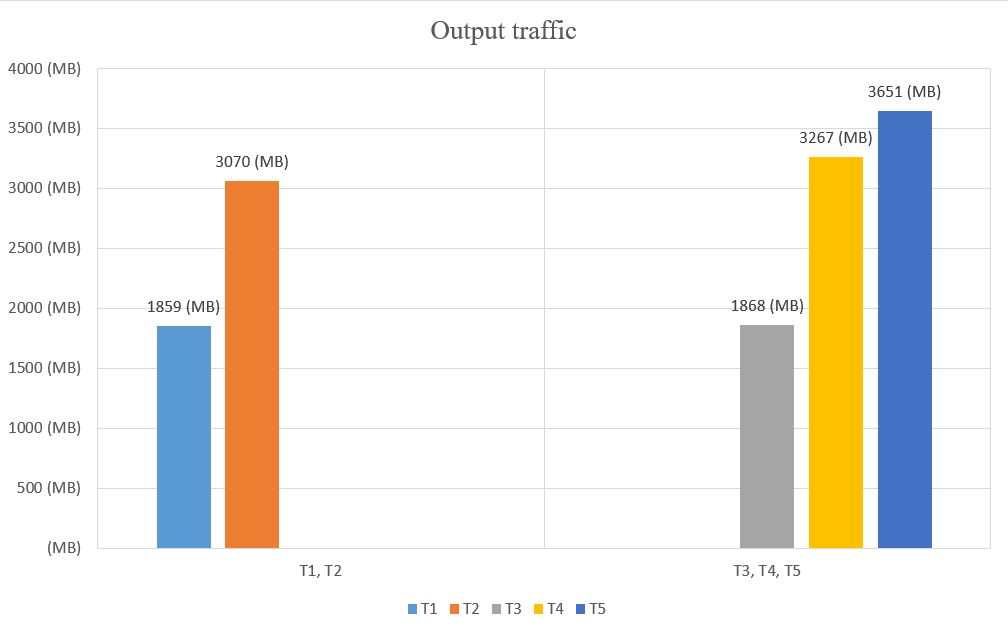}
		\caption{Output traffic}
	\end{subfigure}
	\hfill
	\begin{subfigure}{0.3\textwidth}
		\includegraphics[width=\textwidth]{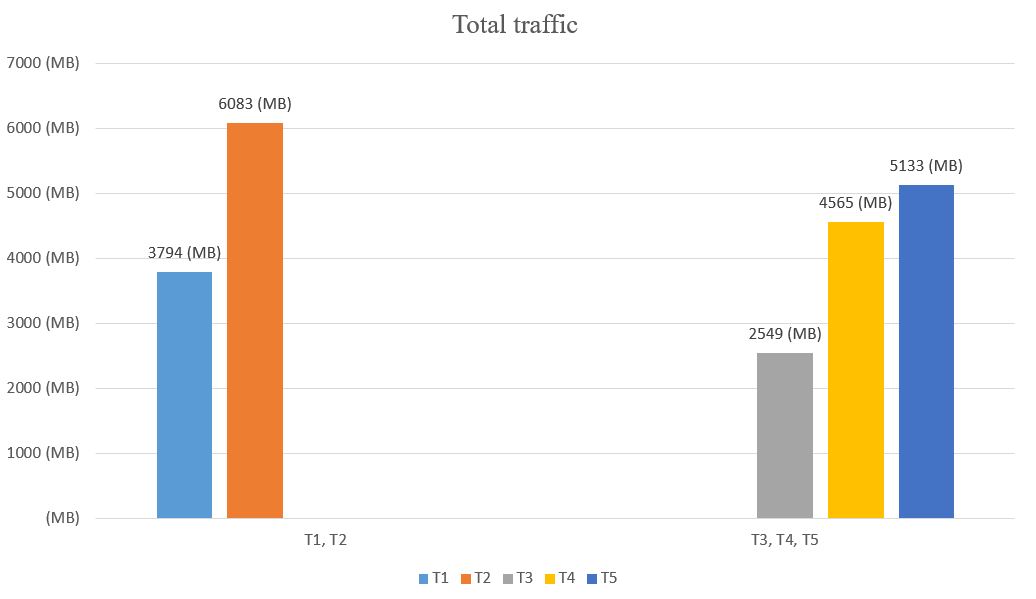}
		\caption{Total traffic }
	\end{subfigure}
	\begin{subfigure}{0.3\textwidth}
		\includegraphics[width=\textwidth]{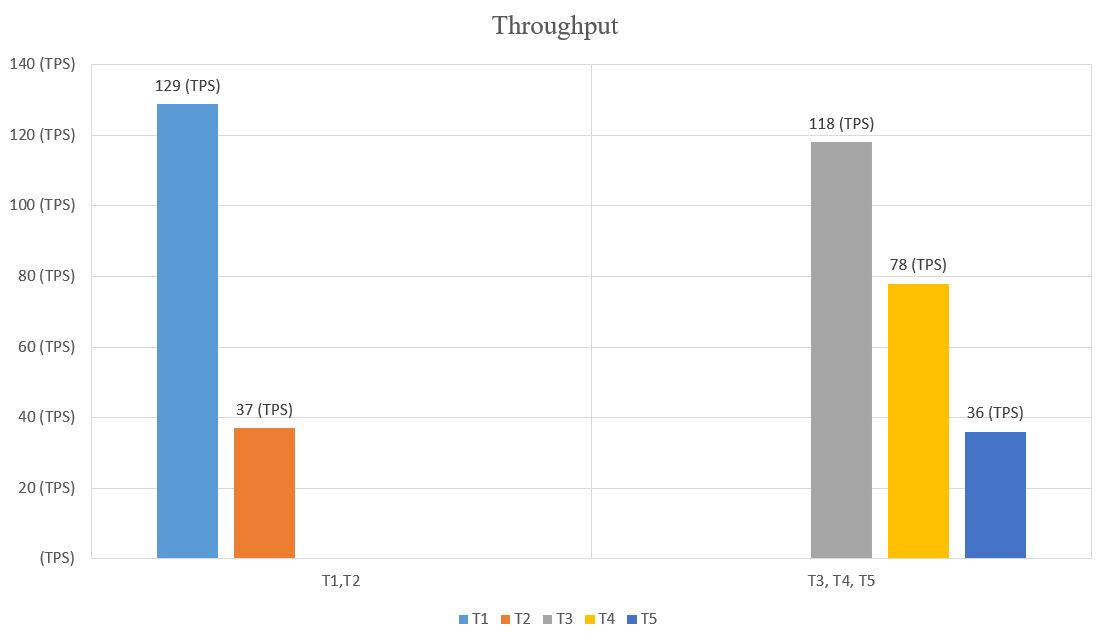}
		\caption{Throughput}
	\end{subfigure}
	\caption{Performance Metrics Evaluation of five scenarios}
	\label{fig:Performance metric evaluation of 5 scenarios}
\end{figure}

The performance of the system was evaluated across five scenarios using parallel and non-parallel configurations, with results captured in Tables \ref{table:Scenario 1} through \ref{table:performance_metrics} and Figure \ref{fig:Performance metric evaluation of 5 scenarios}. Below, we analyze the key metrics and highlight the overall improvements observed in the parallel architecture.
% Compare figures
%table for data

\begin{enumerate}
	\item \textbf{Transaction Throughput (TPS):}  
	As shown in Tables \ref{table:Scenario 1} and \ref{table:Scenario 2}, the parallel architecture achieved a throughput of 129 TPS in Scenario 1 (parallel drug sales registration) compared to 37 TPS in Scenario 2 (non-parallel). Similarly, in Scenario 3 (parallel information retrieval), the system reached 118 TPS, while the non-parallel configuration in Scenario 4 managed only 78 TPS. Across all tested scenarios, the parallel architecture consistently achieved a \textbf{threefold increase in throughput}. This significant boost in throughput demonstrates the clear advantage of parallel processing, particularly in high-demand healthcare environments where timely data access is crucial.
	
	\item \textbf{Resource Utilization (CPU and Memory):}  
	Tables~\ref{table:Scenario 1} through~\ref{table:performance_metrics} indicate that in Scenario 1, the CPU usage of the parallel architecture peaked at 42\%, compared to 32\% in the non-parallel Scenario 2. Memory usage was nearly identical between the two configurations, with the parallel architecture consuming 2429 MiB and the non-parallel configuration consuming 2558 MiB. Thus, while the parallel architecture requires slightly more processing power, its RAM usage remains comparable.

	\item \textbf{Network Traffic (Input/Output Traffic):}  
	The parallel architecture demonstrated significantly reduced network traffic. In Scenario 1, input traffic was 1935 MB, and output traffic was 1859 MB. In contrast, the non-parallel Scenario 2 saw much higher input and output traffic (3013 MB and 3070 MB, respectively). Across all scenarios, network traffic was reduced by \textbf{over 50\%} in the parallel architecture. This reduction is attributed to the distributed nature of parallel transaction processing, which minimizes the need for nodes to handle and propagate all transactions. The reduction in network congestion leads to faster transaction propagation and improved system responsiveness.
\end{enumerate}
The parallel architecture significantly outperforms the non-parallel approach, as shown in Table~\ref{table:performance_metrics}. It achieves 129 TPS in Scenario 1, compared to 37 TPS in Scenario 2, while maintaining similar memory usage (2429 MiB vs. 2558 MiB) and slightly higher CPU usage. Additionally, it reduces network traffic by nearly 50\%, highlighting its scalability and efficiency for healthcare systems requiring high throughput, fast processing, and robust scalability.
\section{Conclusion}\label{sec 5:Conclusion}

This paper has presented a scalable, multi-layered blockchain architecture tailored to address critical challenges in healthcare, particularly in Electronic Health Record (EHR) sharing and drug supply chain management. The architecture's layered design ensures modularity and adaptability, allowing for efficient management of healthcare data while maintaining compliance with privacy regulations such as HIPAA and GDPR. Patient-centric access control empowers individuals to manage their health data securely, while the integration of standardized protocols facilitates interoperability with existing systems. Additionally, the framework enhances transparency and traceability in drug supply chains, offering a reliable method to address counterfeit pharmaceuticals and ensure drug availability. The testing scenarios demonstrated the efficiency of the proposed architecture, particularly in parallel processing configurations, which showed consistent performance improvements over non-parallel setups. These results highlight the framework's ability to handle the data volume and transaction demands of modern healthcare environments, ensuring secure and efficient data processing while maintaining scalability.

\section{References}
[1] A. Adavoudi Jolfaei, S. F. Aghili, and D. Singelee, "A Survey on Blockchain-Based IoMT Systems: Towards Scalability," in IEEE Access, vol. 9, pp. 148948-148975, 2021, doi: 10.1109/ACCESS.2021.3117662. 
\newline
[2] R. Zhang, R. Xue, and L. Liu, "Searchable Encryption for Healthcare Clouds: A Survey," in IEEE Transactions on Services Computing, vol. 11, no. 6, pp. 978-996, Nov.-Dec. 2018, doi: 10.1109/TSC.2017.2762296.
\newline
[3] A. H. Mayer, C. A. da Costa, R. D. R. Righi, "Electronic health records in a Blockchain: A systematic review," Health Informatics J, vol. 26, no. 2, pp. 1273-1288, Jun. 2020, doi: 10.1177/1460458219866350.
\newline
[4] A. A. Sadawi, M. S. Hassan, and M. Ndiaye, "A Survey on the Integration of Blockchain With IoT to Enhance Performance and Eliminate Challenges," in IEEE Access, vol. 9, pp. 54478-54497, 2021, doi: 10.1109/ACCESS.2021.3070555.
\newline
[5] A. M. Eldin, E. Hossny, K. Wassif, and F. A. Omara, "Survey of Blockchain Methodologies in The Healthcare Industry," in 2022 5th International Conference on Computing and Informatics (ICCI), New Cairo, Cairo, Egypt, 2022, pp. 209-215, doi: 10.1109/ICCI54321.2022.9756056.
\newline
[6] P. Pandey, R. Litoriya, "Implementing healthcare services on a large scale: Challenges and remedies based on blockchain technology," Health Policy and Technology, vol. 9, no. 1, pp. 69-78, 2020, doi: 10.1016/j.hlpt.2020.01.004.
\newline
[7] I. Abu-elezz, A. Hassan, A. Nazeemudeen, M. Househ, A. Abd-alrazaq, "The benefits and threats of blockchain technology in healthcare: A scoping review," International Journal of Medical Informatics, vol. 142, pp. 104246, 2020, doi: 10.1016/j.ijmedinf.2020.104246.
\newline
[8] A. A. Mazlan, S. Mohd Daud, S. Mohd Sam, H. Abas, S. Z. Abdul Rasid, and M. F. Yusof, "Scalability Challenges in Healthcare Blockchain System?A Systematic Review," in IEEE Access, vol. 8, pp. 23663-23673, 2020, doi: 10.1109/ACCESS.2020.2969230.
\newline
[9] J. Xie, F. R. Yu, T. Huang, R. Xie, J. Liu, and Y. Liu, "A Survey on the Scalability of Blockchain Systems," in IEEE Network, vol. 33, no. 5, pp. 166-173, Sept.-Oct. 2019, doi: 10.1109/MNET.001.1800290.
\newline
[10] X. Xu, N. Tian, H. Gao, H. Lei, Z. Liu, and Z. Liu, "A Survey on Application of Blockchain Technology in Drug Supply Chain Management," in 2023 IEEE 8th International Conference on Big Data Analytics (ICBDA), Harbin, China, 2023, pp. 62-71, doi: 10.1109/ICBDA57405.2023.10104779.
\newline
[11] F. Hashim, K. Shuaib, F. Sallabi, "MedShard: Electronic Health Record Sharing Using Blockchain Sharding," Sustainability, vol. 13, pp. 5889, 2021. doi: 10.3390/su13115889.
\newline
[12] F.A. Reegu, H. Abas, Y. Gulzar, Q. Xin, A.A. Alwan, A. Jabbari, R.G. Sonkamble, R.A. Dziyauddin, "Blockchain-Based Framework for Interoperable Electronic Health Records for an Improved Healthcare System," Sustainability, vol. 15, pp. 6337, 2023. doi: 10.3390/su15086337.
\newline
[13] A. Diaz, H. Kaschel, "Scalable Electronic Health Record Management System Using a Dual-Channel Blockchain Hyperledger Fabric," Systems, vol. 11, pp. 346, 2023. doi: 10.3390/systems11070346.
\newline
[14] V. Mani and P. M., "ECS Trans. 107, 3233," 2022. doi: 10.1149/10701.3233ecst.
\newline
[15] I.A. Omar, R. Jayaraman, M.S. Debe, K. Salah, I. Yaqoob, and M. Omar, "Automating Procurement Contracts in the Healthcare Supply Chain Using Blockchain Smart Contracts," IEEE Access, vol. 9, pp. 37397-37409, 2021. doi: 10.1109/ACCESS.2021.3062471.
\newline
[16] J. Jayabalan and N. Jeyanthi, "Scalable blockchain model using off-chain IPFS storage for healthcare data security and privacy," Journal of Parallel and Distributed Computing, vol. 164, pp. 152-167, 2022. doi: 10.1016/j.jpdc.2022.03.009.
\newline
[17] K.K.C. Martinez, "Blockchain Scalability Solved via Quintessential Parallel Multiprocessor," in 2023 International Wireless Communications and Mobile Computing (IWCMC), Marrakesh, Morocco, 2023, pp. 1626-1631. doi: 10.1109/IWCMC58020.2023.10183268.
\newline
[18] L. Ismail, H. Materwala, and S. Zeadally, "Lightweight Blockchain for Healthcare," IEEE Access, vol. 7, pp. 149935-149951, 2019. doi: 10.1109/ACCESS.2019.2947613.
\newline
[19] K. Kottursamy, B. Sadayapillai, A.A. AlZubi, A.K. Bashir, "A novel blockchain architecture with mutable block and immutable transactions for enhanced scalability," Sustainable Energy Technologies and Assessments, vol. 58, 103320, 2023. doi: 10.1016/j.seta.2023.103320.
\newline
[20] S. Joshi, A. Choudhury, O. Saraswat, "Enhancing Healthcare System Using Blockchain Smart Contracts," 2022. Available: arXiv:2202.07591.
\newline
[21] Y. Ucbas, A. Eleyan, M. Hammoudeh, and M. Alohaly, "Performance and Scalability Analysis of Ethereum and Hyperledger Fabric," IEEE Access, vol. 11, pp. 67156-67167, 2023. doi: 10.1109/ACCESS.2023.3291618.
\newline
[22] S. Nakamoto, "Bitcoin: A Peer-to-Peer Electronic Cash System," 2008. [Online]. Available: https://bitcoin.org/bitcoin.pdf.
\newline
[23] Z. Zulkifl, et al., "FBASHI: Fuzzy and Blockchain-Based Adaptive Security for Healthcare IoTs," IEEE Access, vol. 10, pp. 15644-15656, 2022, doi: 10.1109/ACCESS.2022.3149046.
\newline
[24] A. N. Gohar, S. A. Abdelmawgoud, and M. S. Farhan, "A Patient-Centric Healthcare Framework Reference Architecture for Better Semantic Interoperability Based on Blockchain, Cloud, and IoT," IEEE Access, vol. 10, pp. 92137-92157, 2022, doi: 10.1109/ACCESS.2022.3202902.
\newline
[25] K. Zala, H. K. Thakkar, R. Jadeja, P. Singh, K. Kotecha, and M. Shukla, "PRMS: Design and Development of Patients? E-Healthcare Records Management System for Privacy Preservation in Third Party Cloud Platforms," IEEE Access, vol. 10, pp. 85777-85791, 2022, doi: 10.1109/ACCESS.2022.3198094.
\newline
[26] A. A. Khan, A. A. Wagan, A. A. Laghari, A. R. Gilal, I. A. Aziz, and B. A. Talpur, "BIoMT: A State-of-the-Art Consortium Serverless Network Architecture for Healthcare System Using Blockchain Smart Contracts," IEEE Access, vol. 10, pp. 78887-78898, 2022, doi: 10.1109/ACCESS.2022.3194195.
\newline
[27] F. Li, K. Liu, L. Zhang, S. Huang, and Q. Wu, "EHRChain: A Blockchain-Based EHR System Using Attribute-Based and Homomorphic Cryptosystem," IEEE Transactions on Services Computing, vol. 15, no. 5, pp. 2755-2765, 2022, doi: 10.1109/TSC.2021.3078119.
\newline
[28] Z. Pang, Y. Yao, Q. Li, X. Zhang, and J. Zhang, "Electronic Health Records Sharing Model Based on Blockchain With Checkable State PBFT Consensus Algorithm," IEEE Access, vol. 10, pp. 87803-87815, 2022, doi: 10.1109/ACCESS.2022.3186682.
\newline
[29] A. R. Rajput, Q. Li, M. T. Ahvanooey, and I. Masood, "EACMS: Emergency Access Control Management System for Personal Health Record Based on Blockchain," IEEE Access, vol. 7, pp. 84304-84317, 2019, doi: 10.1109/ACCESS.2019.2917976.
\newline
[30] X. Liu, Z. Wang, C. Jin, F. Li, and G. Li, "A Blockchain-Based Medical Data Sharing and Protection Scheme," IEEE Access, vol. 7, pp. 118943-118953, 2019, doi: 10.1109/ACCESS.2019.2937685.
\newline
[31] B. U. I. Khan, A. M. Baba, R. F. Olanrewaju, S. A. Lone, and N. F. Zulkurnain, "SSM: Secure-Split-Merge data distribution in cloud infrastructure," in 2015 IEEE Conference on Open Systems (ICOS), Melaka, Malaysia, 2015, pp. 40-45, doi: 10.1109/ICOS.2015.7377275.
\newline
[32] A. P. Singh, et al., "A Novel Patient-Centric Architectural Framework for Blockchain-Enabled Healthcare Applications," IEEE Transactions on Industrial Informatics, vol. 17, no. 8, pp. 5779-5789, 2021, doi: 10.1109/TII.2020.3037889.
\newline
[33] S. Tanwar, K. Parekh, and R. Evans, "Blockchain-based electronic healthcare record system for healthcare 4.0 applications," Journal of Information Security and Applications, vol. 50, 102407, 2020, doi: 10.1016/j.jisa.2019.102407.
\newline
[34] S. Meisami, M. Beheshti-Atashgah, and M. R. Aref, "Using blockchain to achieve decentralized privacy in IoT healthcare," arXiv preprint arXiv:2109.14812, 2021.
\newline
[35] S. Meisami, S. Meisami, M. Yousefi, and M. R. Aref, "Combining Blockchain and IoT for Decentralized Healthcare Data Management," arXiv preprint arXiv:2304.00127, 2023.
\newline
[36] Hyperledger Fabric Samples. GitHub repository. Retrieved from https://github.com/hyperledger/fabric-samples (Accessed Feb 2023).
\newline
[37] R. Javan,"Extend\(_{\text{HLF}}\)testnetwork."GitHub repository. Retrieved from https://github.com/rezajavan/Extend\(_{\text{HLF}}\)testnetwork.
%%=============================================%%
%% For presentation purpose, we have included  %%
%% \bigskip command. Please ignore this.       %%
%%=============================================%%

%%=============================================%%
%% For presentation purpose, we have included  %%
%% \bigskip command. Please ignore this.       %%
%%=============================================%%

\end{document}